\documentclass{article}
\usepackage{times}
\usepackage{graphicx} 
\usepackage{natbib}
\usepackage{algorithm}
\usepackage{algorithmic}
\usepackage {tikz}
\usepackage[font=scriptsize,skip=0pt]{subcaption}

\usetikzlibrary {positioning}
\usepackage{hyperref}
\usepackage{amssymb}
\usepackage{amsmath}
\usepackage{amsthm}

\newcommand{\parallelsum}{\mathbin{\|}}

\makeatletter
\def\thm@space@setup{%
  \thm@preskip=\parskip \thm@postskip=0pt
}
\makeatother

\newtheorem{myremark}{Remark}
\newcommand\independent{\protect\mathpalette{\protect\independenT}{\perp}}
\def\independenT#1#2{\mathrel{\rlap{$#1#2$}\mkern2mu{#1#2}}}

\theoremstyle{plain} 

\theoremstyle{definition}

\theoremstyle{remark}

\usepackage{amsmath,lipsum}
\usepackage{etoolbox}
\newcommand{\zerodisplayskips}{%
\setlength{\abovedisplayskip}{3pt}%
\setlength{\belowdisplayskip}{3pt}%
\setlength{\abovedisplayshortskip}{3pt}%
\setlength{\belowdisplayshortskip}{3pt}}
\appto{\normalsize}{\zerodisplayskips}
\appto{\small}{\zerodisplayskips}
\appto{\footnotesize}{\zerodisplayskips}

\usepackage[accepted]{icml2017}
\icmltitlerunning{A Unified Bellman Equation for Causal Information and Value in Markov Decision Processes}
\begin{document} 

\captionsetup[figure]{font=small,skip=0pt}

\setlength{\parskip}{0pt}
\setlength{\abovedisplayskip}{0pt}
\setlength{\belowdisplayskip}{0pt}
\setlength{\abovedisplayshortskip}{0pt}
\setlength{\belowdisplayshortskip}{0pt}

\twocolumn[
\icmltitle{A Unified Bellman Equation for Causal Information and Value \\in Markov Decision Processes}
\icmlsetsymbol{equal}{*}

\begin{icmlauthorlist}
\icmlauthor{Stas Tiomkin}{cse}
\icmlauthor{Naftali Tishby}{cse,elsc}
\end{icmlauthorlist}

\icmlaffiliation{elsc}{The Edmond and Lilly Safra Center for Brain Sciences, The Hebrew University, Jerusalem, Israel}
\icmlaffiliation{cse}{The Benin School of Computer Science and Engineering, The Hebrew University,  Jerusalem, Israel.}

\icmlcorrespondingauthor{Stas Tiomkin}{stas.tiomkin@mail.huji.ac.il}
\icmlcorrespondingauthor{Naftali Tishby}{tishby@cs.huji.ac.il}

\icmlkeywords{ Reinforcement learning, causal information, Bellman equation, Lyapunov function}
\vskip 0.3in
]
\printAffiliationsAndNotice{}  
\begin{abstract} 
The interaction between an artificial agent and its environment is bi-directional. The agent extracts relevant information from the environment, and affects the environment by its actions in return to accumulate high expected reward. Standard reinforcement learning (RL) deals with the expected reward maximization. However, there are always information-theoretic limitations that restrict the expected reward, which are not properly considered by the standard RL.

In this work we consider RL objectives with information-theoretic limitations. For the first time we derive a Bellman-type recursive equation for the causal information between the environment and the agent, which is combined plausibly with the Bellman recursion for the value function. The unified equitation serves to explore the typical behavior of artificial agents in an infinite time horizon. 

\end{abstract} 

\section{Introduction}
The interaction between an organism and the environment consists of three major components - utilization of past experience, observation of the current environmental state, and generation of a behavioral policy. The latter component, {\it planning} is an essential feature of intelligent systems for survival in limited-resource environments. Organisms with long-term planning can accumulate more resources and avoid possible catastrophic states in the future, which makes than evolutionarily superior to 'short-term planners'. 

An intelligent agent combines past experience with the current observations to act upon the environment. This feedback interaction induces complex statistical correlations over different time scales between environment state trajectories and the agent's action trajectories. Infinite-time interaction patterns, (infinite state and action trajectories) define the typical behavior of an organism in a given environment. This typical behavior is crucial for the design and analysis of intelligent systems. In this work we derive typical behavior within the formalism of a reinforcement learning (RL) model subject to information-theoretic constraints.
  
In the standard RL model, \cite{sutton1998reinforcement}, an artificial organism generates an optimal policy through an interaction with the environment. Typically, the optimality is taken with regard to a reward, such as energy, money, social network 'likes/dislikes', time etc. Intriguingly, the reward-associated value function, (an average accumulated reward), possesses the same property for different types of rewards - the optimal value function is a {\it Lyapunov function}, \cite{perkins2002lyapunov}, a {\it generalized energy} function. 

A principled way to solve RL (as well as Optimal Control) problems is to find a corresponding generalized energy function. Specifically, the Bellman recursive equation, \cite{bertsekas1995dynamic}, is the {\it Lyapunov function} for Markov decision processes. Strictly speaking, the standard RL framework is about the discovery and minimization (maximization) of an {\it energetic} quantity, whereas the Bellman recursion is an elegant tool for finding the optimal policy both for finite and infinite time horizons for planning.   

However, this model is incomplete. Physics suggests that there is always an interplay between energetic and entropic quantities. Recent advances in the artificial intelligence show that the behavior of {\it realistic} artificial agents is indeed affected simultaneously by energetic and entropic quantities \cite{tishby2011information, rubin2012trading}. The finite rate of information transfer is a fundamental constraint for any value accumulation mechanism. 

Information-theoretic constraints are ubiquitous in the interaction between an agent and the environment. For example, a sensor has a limited bandwidth, a processor has a limited information processing rate, a memory has a limited information update rate, etc. All these are not pure theoretical considerations, but rather the practical requirements to build and analyze realistic intelligent agents. Moreover, often, an artificial agent needs to limit the available information to the information for a particular task, \cite{polani2006relevant}. For example, when looking at a picture, the visual stimuli strikes the retina at the rate of the speed of light, which is decreased dramatically to leave only the relevant information rate, which is essential for understanding the picture.   

Different types of entropic constraints have been studied in the context of RL. The $KL$-control, \cite{todorov2006linearly, tishby2011information,kappen2012optimal} introduces a $D_{KL}$ cost to each state, which penalizes complex policies, where complexity is measured by the $D_{KL}$ divergence between a prior policy, or uncontrolled dynamics, and an actual policy. Importantly, the $KL$-control does not directly address the question of the information transfer rate. 

Some recent works do explicitly consider the effects of information constraints on the performance of a controller \cite{tatikonda2004control, tanaka2015sdp, tanaka2015lqg} for an infinite time horizon. However, these works are mostly limited to linear dynamics. 

Obviously, feedback implies causality - the current action depends on previous and/or present information alone. Consequently, the information constraints should address the interaction causality.     

In this work we consider directed information constraints over an infinite time horizon in the framework of Markov decision processes, where the directed information is between the environment state trajectories and the agent's action trajectories. 

For the first time we derive a Bellman type equation for the causal information, and combine it with the Bellman recursion for value. The unified Information-Value Bellman equation enables us to derive the typical optimal behavior over an infinite time horizon in the MDP setting. 

In addition, this work has practical implications for the design criteria of optimal intelligent agents. Specifically, an information processing rate of a 'brain' (processor) of an artificial agent should be higher than the minimal information rate required to solve a particular MDP problem.

The interaction between an agent and the environment is bi-directional - an agent extracts information and affects the environment by its actions in return. To provide a comprehensive analysis of this bi-directional interaction, we consider both information channels. These channels are dual, comprising the action-perception cycle of reinforcement learning under the information-theoretic constraints. We derive a Bellman type equation for the directed information from action trajectories to state trajectories. 

The minimal directed information rate from the environment to the agent comprises a constraint for the expected reward maximization in standard goal-oriented RL test bench tasks. By contrast, the maximal directed information rate from the agent to the environment comprises a {\it proxy} for solving different AI test bench tasks without specifying the target, which can be considered as self-motivated behavior to be ready for any possible target. We elaborate on the difference between these two opposite information rates by their corresponding analytic expressions, and by numerical simulations of different types of problems, which can be solved by them.            

The paper is organized as follows. In Section \ref{sec:Prelim} we provide background on causal conditioning, directed information, and Markov decision processes. In Section \ref{sec:DIagent2evn} we represent the Bellman-type recursive equation for directed information from the agent's actions to the environment. In Section \ref{sec:UnifiedBellman1} we show the unified recursion for the directed information and the common recursion for the value function. The optimization problem for finding the minimal directed information rate required to achieve the maximal expected reward in an infinite time horizon is stated in Section \ref{sec:optProb1}. The numerical simulation of the unified Bellman equation is provided in Section \ref{sec:numSim1}, where we consider the standard maze-escaping problem for a predefined target. In Section \ref{sec:dualProb} we consider the dual problem of the maximal directed information rate from the agent to the environment, and derive the dual Bellman equation. In Section \ref{sec:UnknownGoal} we provide a numerical simulation for the dual Bellman equation, where we consider tasks without a predefined target. We compare our solution to the standard algorithms for finding the average all-pairs shortest path problem. Finally, in Section \ref{sec:disc} we summarize the paper, and provide directions for the continuation of this work.

\section{Background}\label{sec:Prelim}

In this section we overview the theoretical background for this work. Specifically, we briefly review the framework of reinforcement learning and Markov decision processes. Then, we review causal conditioning and directed information, and define the required quantities.

\subsection{Interaction Model}\label{sec:InterModel}

\begin{figure}[h!]
\begin {center}
\begin{tikzpicture}[scale=.6, transform shape, every node/.style={thick,circle,inner sep=0pt}]
\node[circle, fill=green!40, minimum size=1cm] (a2) at (4, 0)  {$A_{\infty}$};
\node[circle, fill=green!40, minimum size=1cm] (a3) at (6, 0)  {$A_{t-1}$};
\node[circle, fill=green!40, minimum size=1cm] (a4) at (8, 0)  {$A_{t}$};
\node[circle, fill=green!40, minimum size=1cm] (a5) at (10, 0) {$A_{t+1}$};
\node[circle, fill=green!40, minimum size=1cm] (a6) at (12, 0) {$A_{T-1}$};
\node[circle, fill=green!40, minimum size=1cm] (a7) at (14, 0) {$A_{T}$};
\node[circle, fill=red!60, minimum size=1cm] (x2) at (3, 1.65) {$S_{\infty}$};
\node[circle, fill=red!60, minimum size=1cm] (x3) at (5, 1.65) {$S_{t-1}$};
\node[circle, fill=red!60, minimum size=1cm] (x4) at (7, 1.65) {$S_t$};
\node[circle, fill=red!60, minimum size=1cm] (x5) at (9, 1.65) {$S_{t+1}$};
\node[circle, fill=red!60, minimum size=1cm] (x6) at (11, 1.65) {$S_{T-1}$};
\node[circle, fill=red!60, minimum size=1cm] (x7) at (13, 1.65) {$S_{T}$};
\foreach \from/\to in { x3/a3, x3/x4, a3/x4, x4/x5, x4/a4, a4/x5, x5/a5, x6/a6, x6/x7, a6/x7, x7/a7}
\draw [->] (\from) -- (\to);
\foreach \from/\to in {x2/a2, x2/x3,a2/x3,x5/x6,a5/x6}
\draw [dotted] (\from) -- (\to);
\end{tikzpicture}
\end{center}
\caption{A probabilistic directed graph for a Markov Decision Process.}\label{MDP_graph}
\end{figure}
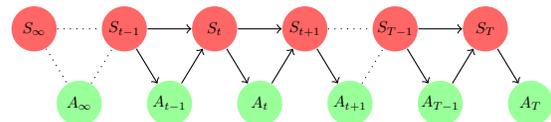

We assume that the agent and environment interact at discrete time steps for $t\in (-\infty, \dots, T)$. At time $t$, the environment appears at state $S_t\in \mathcal{S}$, the agent observes the state, and affects the environment by its action $A_t\in \mathcal{A}$ according to the policy $\pi(A_t\mid S_t)$. The environmental evolution is given by the transition probability distribution, $p(S_{t+1}\mid A_t, S_t)$. This interaction model is plausibly described by the probabilistic graph, shown at Figure \ref{MDP_graph}, where the arrows denote directions of causal dependency.

As a results of this feedback interaction, the environment state trajectories, $S^T_{-\infty}$ defined by (\ref{eq:StateTraj}), and the agent action trajectories, $A^T_{-\infty}$ defined by (\ref{eq:ActionTraj}), become statistically correlated.

{\small
\begin{align}
S^T_{-\infty} \doteq& (\dots, S_{t-1}, S_t, S_{t+1}, \dots, S_T),\label{eq:StateTraj}\\
A^T_{-\infty} \doteq& (\dots, A_{t-1}, A_t, A_{t+1}, \dots, A_T).\label{eq:ActionTraj}.
\end{align}
}

The corresponding trajectories of finite length length $T$, and their time-delayed versions are given by:

{\small
\begin{align}
S^T_{t+1} \doteq& (S_{t+1}, S_{t+2}, \dots, S_T),\nonumber\\
A^T_{t+1} \doteq& (A_{t+1}, A_{t+2}, \dots, A_T),\nonumber\\
DS^T_{t+1} \doteq& (S_{t}, S_{t+1}, \dots, S_{T-1}),\nonumber\\
DA^T_{t+1} \doteq& (A_{t}, A_{t+1}, \dots, A_{T-1}),\nonumber
\end{align}
}

where $DS$ denotes the time-delay operator.
\begin{myremark}\label{re:Intersection} 
These definitions imply that the original sequence, $S^T_{t+1}$, includes the time-delayed sequence, $DS^T_{t+1}$, except for its first component, $S_t$. The union between $S^T_{t+1}$ and $DS^T_{t+1}$ is a sequence $S_t^T$ with length $t+1$. 
\end{myremark}

\subsubsection{Bellman Recursion for Value}
Extending the interaction model in Section \ref{sec:InterModel} by introducing the reward function, $r(S_{t+1}\mid A_t, S_t)$, makes the agent to plan 'consciously' its actions by adapting the policy, $\pi(A_t\mid S_t)$, in order to achieve a higher reward in a given time-horizon. The optimal policy, $\pi(A_t\mid S_t)$, that guarantees the maximal expected reward, is found by the {\it policy iteration algorithm}, \cite{sutton1998reinforcement}, where a value of $\pi(A_t\mid S_t)$ is estimated by the value function. It can be computed recursively: 

{\small
\begin{align}
Q^{\pi}(S_{t}, A_{t}) =& \Bigl<r({S_{t+1}; A_t, S_t})\Bigr>_{p(S_{t+1}\mid S_{t}, A_t)}\label{eg:ValREC}\\
+& \gamma_t\Bigl<Q(S_{t+1}, A_{t+1})\Bigr>_{\pi(A_{t+1}\mid S_{t+1})p(S_{t+1}\mid S_{t}, A_t)},\nonumber
\end{align}
}

where, $\gamma_t$ is the discounting factor. The Bellman recursion is an fundamental concept in optimal planning. A comprehensive review of reinforcement learning can be found in \cite{bertsekas1995dynamic, sutton1998reinforcement,puterman2014markov}.  

\subsection{Causal Conditioning}

A random process can be conditioned by another random process either causally or non-causally, respectively:

{\small
\begin{align}
p(A_{t+1}^T\parallelsum S_{t+1}^T) \doteq& \prod\limits_{i=t+1}^Tp(A_i\mid A_{t+1}^{i-1}, \mathbf{S_{t+1}^i})\label{causCondAS}\\
p(A_{t+1}^T\mid S_{t+1}^T) \doteq& \prod\limits_{i=t+1}^Tp(A_i\mid A_{t+1}^{i-1}, \mathbf{S_{t+1}^T}).
\end{align}
}

The causal conditioning assumes {\it time-synchronization} between the sequences.  
A random process can be causally-conditioned by a random variable at a particular time as well. Compare the following causal 

{\small
\begin{align}
p(A_{t+1}^T\parallelsum S_{t+\tau})\! =& \!\!\!\!\prod\limits_{i=t+1}^{t+\tau-1}\!\!\!p(A_i\mid A_{t+1}^{i-1})\nonumber
					\!\!\prod\limits_{i=t+\tau}^{T}\!\!p(A_i\mid A_{t+1}^{i-1}, S_{t+\tau}),
\end{align}
} 

vs. non-causal scalar conditioning:

{\small
\begin{align}
p(A_{t+1}^T\mid S_{t+\tau}) =& \prod\limits_{i=t+1}^{T}p(A_i\mid A_{t+1}^{i-1}, S_{t+\tau})\nonumber.
\end{align}
} 

A random sequence can be conditioned simultaneously by another random sequence and by a scalar random variable. In particular, the following two expressions will be useful in the next section:  

{\small
\begin{align}
p(A_{t+1}^T\parallelsum S_{t}^T, A_t) =& \prod\limits_{i=t+1}^Tp(A_i\mid S_{t}^i, A_{t+1}^{i-1}, A_t)\nonumber\\
p(A_{t+1}^T \parallelsum S_{t}^{T-1}, A_t)=& \prod\limits_{i=t+1}^Tp(A_i\mid S_{t}^{i-1}, A_{t+1}^{i-1}, A_t),\nonumber
\end{align}
}
\normalsize{which, due to the graph in Figure \ref{MDP_graph}, are reduced to:}
{\small
\begin{align}
p(A_{t+1}^T\parallelsum S_{t}^T, A_t) =& \prod\limits_{i=t+1}^Tp(A_i\mid S_i)\label{DIpNum}\\
p(A_{t+1}^T \parallelsum S_{t}^{T-1}, A_t)=& \prod\limits_{i=t+1}^Tp(A_i\mid S_{i-1}, A_{i-1})\label{DIpDen}.
\end{align}
}
\subsubsection{Causal Entropy}
Following the causally conditioned distribution (\ref{causCondAS}), the causally conditioned entropy is defined by:  

{\small
\begin{align}
	H(A_{t+1}^T\parallelsum S_{t+1}^T) =& \Bigl<\ln p(A_{t+1}^T\parallelsum S_{t+1}^T)\Bigr>_{p(A_{t+1}^T, S_{t+1}^T)}.
\end{align}
}
The causally conditioned entropy satisfies the chain rule, (and other properties of entropy, \cite{kramer1998directed}): 

{\small
\begin{align}
H(A_{t+1}^T\parallelsum S_{t+1}^T)\sum\limits_{i=t+1}^TH(A_i\mid S_{t+1}^i, A_{t+1}^{i-1}).
\end{align}
}

The causally conditioned entropy enables us to define the directed (causal) information between two sequences of random variables. 

\subsubsection{Directed Information}
The Shannon information between two sequences of random variables, $X_1^T$, and, $Y_1^T$, is given by a decrease in the entropy of one variable by the conditional entropy of this variable \cite{cover2012elements}:

{\small
\begin{align}
\mathcal{I}\Bigl[X;Y\Bigr] \doteq H(X) - H(X\mid Y).\label{eq:MIORD}
\end{align}
}

The directed information is defined similarly as a decrease in the entropy. But, in this case, the conditional entropy is the causally conditional entropy \cite{massey2005conservation, kramer1998directed}:

{\small
\begin{align}
\vec{\mathcal{I}}\Bigl[X\rightarrow Y\Bigr] \doteq H(X) - H(X \parallelsum Y).\label{eq:MIDI}
\end{align}
}

The essential difference between ordinary mutual information (\ref{eq:MIORD}) and directed information (\ref{eq:MIDI}) is the lack of symmetry:

{\small
\begin{align}
\vec{\mathcal{I}}\Bigl[X\rightarrow Y\Bigr] \neq \vec{\mathcal{I}}\Bigl[Y\rightarrow X\Bigr]. 
\end{align}
}

A comprehensive review of directed information properties can be found in \cite{kramer1998directed, marko1973bidirectional, massey2005conservation}.

\section{ Information from Environment to Agent }\label{sec:DIagent2evn}
We equip the agent with the ability to remember the previous state and the action. Memory is a desired feature of intelligent agents. Memoryless agents can act only reactively without learning from the experience. 
Specifically, we model the memory by {\it causally-conditioning} the directed information flow from the environments to the agent. Assuming that at time $t$ the environment and the agent action were $S_t$ and $A_t$, respectively, the causally-conditioning directed information is given by: 

{\small
\begin{align}
\vec{\mathcal{I}}\Bigl[&S_{t+1}^T\rightarrow A_{t+1}^T\parallelsum DS_{t+1}^T, A_t\Bigr]=\nonumber\\
=&\Bigl<\ln\Bigl[\frac{p(A_{t+1}^T\parallelsum S_{t+1}^T, DS_{t+1}^T, A_t)}{ p(A_{t+1}^T\parallelsum DS_{t+1}^T, A_t) }\Bigr]\Bigr>_{p(S_{t+1}^T, A_{t+1}^T\mid S_t, A_t)},\nonumber\\
\intertext{\normalsize{which, following Remark \ref{re:Intersection}, is equivalent to:}}
=&\Bigl<\ln\Bigl[\frac{p(A_{t+1}^T\parallelsum S_{t}^T, A_t)}{ p(A_{t+1}^T \parallelsum S_{t}^{T-1}, A_t) }\Bigr]\Bigr>_{p(S_{t+1}^T, A_{t+1}^T\mid S_t, A_t)}\label{eq:DI1},\\
\intertext{\normalsize{and, according to Eq. (\ref{DIpNum}) and Eq. (\ref{DIpDen}) is decomposed to:}}
=&\Bigl<\ln\Bigl[\frac{ \prod\limits_{i=t+1}^Tp(A_i\mid S_i)  }{ \prod\limits_{i=t+1}^Tp(A_i\mid S_{i-1}, A_{i-1})  }\Bigr]\Bigr>_{p(S_{t+1}^T, A_{t+1}^T\mid S_t, A_t)}\nonumber\\
=&  \sum\limits_{i=t+1}^T  \Bigl<\ln\Bigl[\frac{ p(A_i\mid S_i)  }{ p(A_i\mid S_{i-1}, A_{i-1})  }\Bigr]\Bigr>_{p(S_{t+1}^T, A_{t+1}^T\mid S_t, A_t)}\label{eq:DI2}.
\end{align}
}  

\subsection{Bellman Equation for $\vec{\mathcal{I}}[\mbox{States} \rightarrow \mbox{Actions}]$ } 
The expression (\ref{eq:DI2}) enables us to write the directed information recursively:

{\small
\begin{align}
\vec{\mathcal{I}}\Bigl[S_{t+1}^T&\rightarrow A_{t+1}^T\parallelsum DS_{t+1}^T, A_t\Bigr]=\nonumber\\
=&\Bigl<\ln\Bigl[\frac{ p(A_{t+1}\mid S_{t+1})  }{ p(A_{t+1}\mid S_{t}, A_{t})  }\Bigr]\Bigr>_{p(S_{t+1}, A_{t+1}\mid S_{t}, A_{t})}\nonumber\\
+&  \sum\limits_{i=t+2}^T  \Bigl<\ln\Bigl[\frac{ p(A_i\mid S_i)  }{ p(A_i\mid S_{i-1}, A_{i-1})  }\Bigr]\Bigr>_{p( S_{t+1}^T, A_{t+1}^T\mid S_t, A_t )}\nonumber\\
=& {\mathcal{I}}\left[S_{t+1}; A_{t+1}\mid S_{t}, A_{t}\right]\label{eq:DI3}\\
+& \Bigl< \vec{\mathcal{I}}\Bigl[S_{t+2}^T\!\!\rightarrow\!\!A_{t+2}^T\!\parallelsum\! DS_{t+2}^T, \!A_{t+1}\Bigr]\Bigr>_{ p(S_{t+1}, A_{t+1}\mid S_{t}, A_{t})}\nonumber. 
\end{align}
} 

The directed information in (\ref{eq:DI1}) is a function of $S_t$ and $A_t$, whereas the directed information in (\ref{eq:DI3}) within the expectation operator $<\cdot>_{  p(S_{t+1}, A_{t+1} \mid S_{t}, A_{t})   }$ is a function of $S_{t+1}$ and $A_{t+1}$. Consequently, the recursive equation in (\ref{eq:DI3}) can be decomposed into the first step and all the future steps:   

{\small
\begin{align}
\vec{\mathcal{I}}_T(S_t,& A_t) = {\mathcal{I}}\left[S_{t+1}; A_{t+1}\mid S_{t}, A_{t}\right]\label{eq:DIREC}\\
+& \Bigl< \vec{\mathcal{I}}_T(S_{t+1}, A_{t+1})\Bigr>_{\pi(A_{t+1}\mid S_{t+1})p(S_{t+1}\mid S_{t}, A_{t})}\nonumber,
\end{align}
}

where the first term is the local step of the recursion and the second term is the future directed information. This expression comprises the Bellman type equation for the directed information between the environment and the agent. The time horizon of the future state and action trajectories is marked by a subscript in the directed information, $\vec{\mathcal{I}}_T(S_t, A_t)$.

Information processing is an integral part of the interaction between the agent and its environment. An optimal agent strives to accumulate as much value as possible under the constraints of limited bandwidth and information transfer rate. 
In the next section we show that the Bellman recursion for the directed information and the Bellman recursion for the value naturally combine into a unified Bellman recursion. 

\section{Unified Bellman Recursion}\label{sec:UnifiedBellman1}

As mentioned above, the optimal control solution should adhere to the information processing constraints, which are always present in real-life intelligent systems. For example, the limited bandwidth and/or time-delays due to the distance between a controller and a plant are ubiquitous information-theoretic constraints, which may reduce the maximal achievable value. 

Consequently, the designer of an intelligent agent should always address the question -  {\it what is the minimal information rate required to achieve the maximal value?} This question is formulated by the unconstrained Lagrangian which comprises the unified Bellman recursion. 

We found that the Bellman recursion for the directed information (\ref{eq:DIREC}) and the Bellman recursion for the value (\ref{eg:ValREC}) combine naturally to form the unified {\it Information-Value Bellman Recursion}:

{\small
\begin{align}
\mathcal{G}^{\pi}_T(S_t, A_t, \beta) =& \vec{\mathcal{I}}_T^{\pi}(S_t, A_t) - \beta Q^{\pi}(S_{t}, A_{t})\nonumber=
\end{align}
}

which is decomposed into the first step and all the future steps: 

{\small
\begin{align}
= {\mathcal{I}}^{\pi}_T\Bigl[&S_{t+1}; A_{t+1}\mid S_{t}, A_{t}\Bigr] \label{eq:GREC} \\
&\qquad\quad - \beta \left<r(S_{t+1}; S_t, A_t)\right>_{p(S_{t+1}\mid S_t, A_t)}\nonumber\\
+& \Bigl< \vec{\mathcal{I}}_T^{\pi}(S_{t+1}, A_{t+1}) \nonumber\\ 
&\qquad\quad-\!\!\beta Q^{\pi}\!(S_{t+1},\!A_{t+1})\!\Bigr>_{\!\!\pi(A_{t+1}\mid S_{t+1})p(S_{t+1}\mid S_{t}, A_{t})}\nonumber\!\!.
\end{align}
}

Writing explicitly the mutual information and identifying the terms within the second expectation operator with the future averaged value of $\mathcal{G}^{\pi}_T(S_{t+1}, A_{t+1}, \beta)$, we get the unified Bellman recursion:

{\small
\begin{align}
\mathcal{G}^{\pi}_T(S_t, A_t, \beta) =& \Bigl<\pi(A_{t+1}\mid S_{t+1})\ln\Bigl[\frac{\pi(A_{t+1}\mid S_{t+1})}{p(A_{t+1}\mid S_{t}, A_{t})}\Bigr]\label{eq:GREC2}\\
&\qquad\qquad- \beta r(S_{t+1}; S_t, A_t)\Bigr>_{p(S_{t+1}\mid S_t, A_t)}\nonumber\\
+& \Bigl<\mathcal{G}^{\pi}_T(S_{t+1}, A_{t+1},\beta)\Bigr>_{\pi(A_{t+1}\mid S_{t+1})p(S_{t+1}\mid S_{t}, A_{t})}\!\!.\nonumber
\end{align}
} 

This recursive equation can be computed for any $T$. Specifically, we are interested in typical behaviour, when an agents acts according to a stationary policy, $\pi(A\mid S)$. In the next section we show how to find the optimal policy, $\pi(A\mid S)$, that maximizes the unified Bellman recursion in (\ref{eq:GREC2}). 

\subsection{Optimization Problem}\label{sec:optProb1}
To find the optimal policy (that minimizes the directed information under a constraint on the attained value), one needs to solve the unconstrained minimization of the corresponding Lagrangian in (\ref{eq:GREC}).

{\small
\begin{align}
\underset{\pi}{\mbox{argmin}}\mathcal{G}_T^{\pi}(S_t,\!A_t,\!\beta)\!\!=\!\!\underset{\pi}{\mbox{argmin}}\Bigl[\vec{\mathcal{I}}^{\pi}_T(S_t,\!A_t)\!\!-\!\!\beta Q^{\pi}(S_{t},\!A_{t})\Bigr]\label{eq:OPTPR}.
\end{align}
}

The minimization with regard to $\pi$ is done by augmenting the equation in (\ref{eq:OPTPR}) by the Lagrange term for the normalization of $\pi(A\mid S)$, and computing the variational derivative, ${\delta \mathcal{G}_T^{\pi}}/{\delta \pi}$. The minimization with regard to $q(A_{t^{'}}\mid S, A)\doteq p(A_{t^{'}}\mid S, A)$ is done by setting the corresponding variational derivative ${\delta \mathcal{G}_T^{q}}/{\delta q}$ to zero. Straightforward computation of these variational derivatives gives the following set of self-consistent equations:  

{\small
\begin{align}
&\pi^*(A\mid S) = \frac{\exp\left(-\mathcal{G}^{\pi^*,q^*}(S_t, A_t, \beta)\right)}{\sum\limits_{A}\exp\left(-\mathcal{G}^{\pi^*, q^*}(S_t, A_t, \beta)\right)},\label{eq:OPTPI}\\
&p^*(A^{'}\mid S, A) = \frac{1}{\left|\mathcal{A}\right|},\label{eq:OPTQ}
\end{align}
} 
  
\subsection{Stationary Policy}
The typical behavior of the agent in its environment is defined  by the stationary distribution over the states and the actions. This stationary distribution is given by the policy with the infinite time horizon:

{\small
\begin{align}
\pi^*(A\mid S) = \frac{\exp\left(-\mathcal{G}^{\pi^*,q^*}(S, A, \beta)\right)}{\sum\limits_{A}\exp\left(-\mathcal{G}^{\pi^*, q^*}(S, A, \beta)\right)}.
\end{align}
} 

The unified Bellman recursion enables us to compute the stationary policy by iterating the recursion in Eq. (\ref{eq:GREC2}) in the same way as for the Bellman recursion for the value function. Previously, the stationary policy for directed information minimization was derived for the LQG setting \cite{tanaka2015lqg}; in this work we derive the optimal stationary policy the MDP setting. 

\section{Numerical Simulations}\label{sec:numSim1}
To explore a long-term (typical) interaction between an agent and the environment we solve different benchmark RL problems, such as {\it maze escaping problems} \cite{sutton1998reinforcement}, using the information-value recursion in Eq. (\ref{eq:GREC2}). 

We emphasize that even though, in these experiments the target goal appears in a particular state (or at a number of states), there are different possible scenarios where a goal state can appear in any state. In these cases, the agent should be prepared to reach any goal. This {omni-goal} strategy will be discussed in Section \ref{sec:UnknownGoal}. 

\subsection{Maze with predefined targets}\label{sec:MazeWithPredefined}
To investigate the solutions in (\ref{eq:GREC2}) we simulated different mazes where the agent is asked to find a target cell. The agent has full information about the world (it knows its exact location within the maze), which is proportional to the logarithm of the maze size. But it cannot utilize all this information to generate the optimal actions due to its slow 'brain'. For example, the agent's sensor retrieves $8$ bits of information about its location in each interaction with the environment, whereas the processor can transform only $3$ bits of information to the policy. 

The dynamics is given by a transition probability $p(S_{t+1}\mid A_t, S_t)$. If a certain action at time $t$ is impossible, such as stepping into a wall, the agent remains in its current position. The target state is an absorbing state - the agent stays in this state for any action $p(S_{{target}}\mid A_t, S_t)=\delta_{S_{{target}},S_t}$. The agent receives a reward, $0$, when it is in the target state. In all other states the reward is given by $r(S_{t+1}\mid A_t, S_t)=-1$. This is the general setting for maze escaping problems, \cite{sutton1998reinforcement}. 

\begin{figure}[t!]
\begin{center}
\begin{subfigure}{0.49\linewidth}
\includegraphics[width=\linewidth]{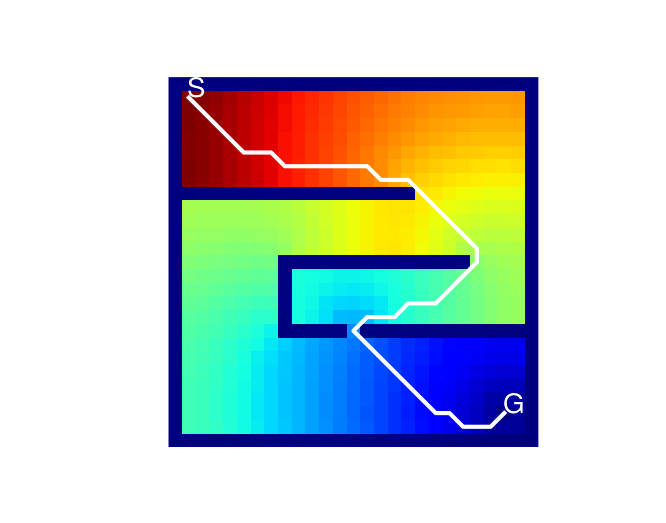}
\caption{}
\end{subfigure}
\begin{subfigure}{0.49\linewidth}
\includegraphics[width=\linewidth]{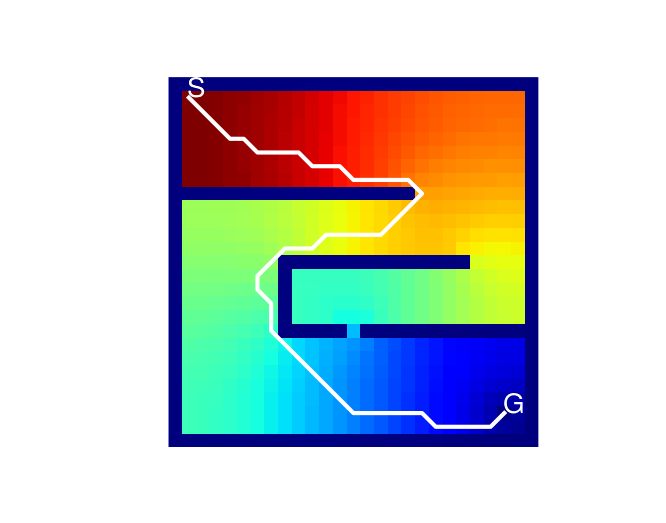}
\caption{}
\end{subfigure}

\begin{subfigure}{0.49\linewidth}
\includegraphics[width=\linewidth]{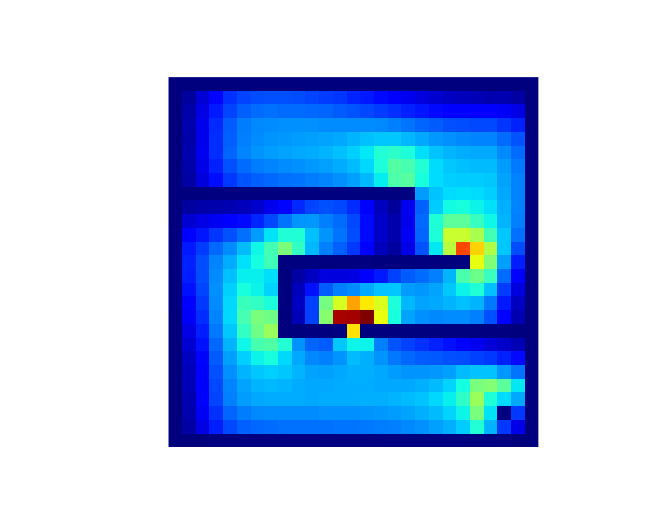}
\caption{}
\end{subfigure}
\begin{subfigure}{0.49\linewidth}
\includegraphics[width=\linewidth]{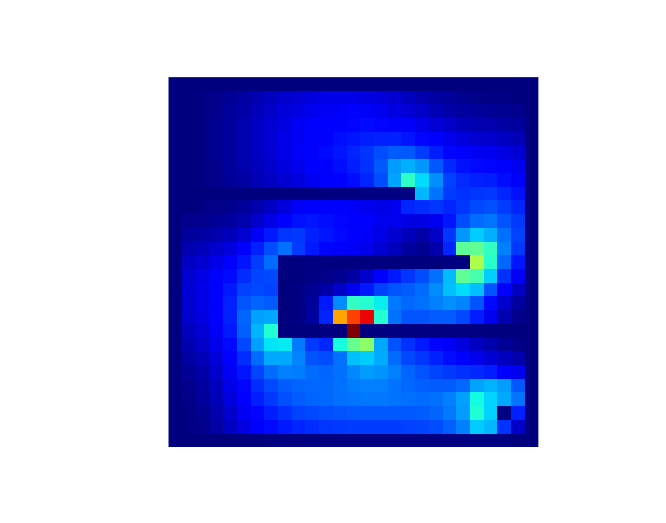}
\caption{}
\end{subfigure}\hspace{\fill}

\begin{subfigure}{0.49\linewidth}
\includegraphics[scale=0.175]{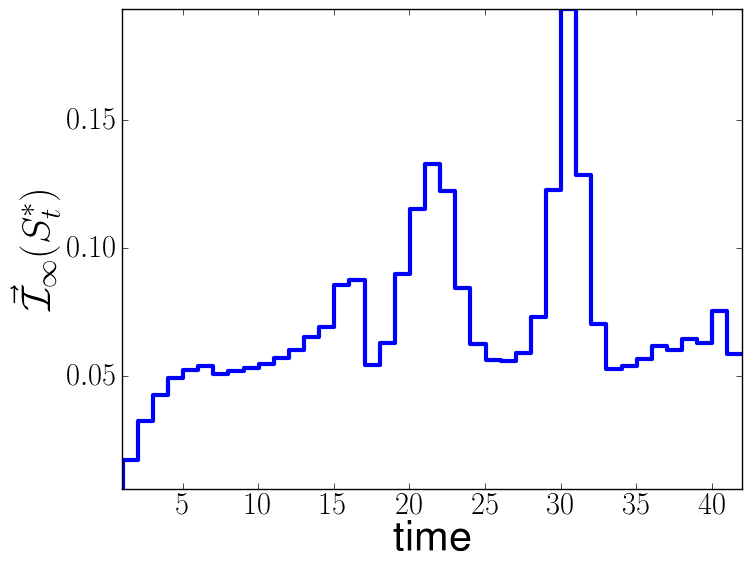}
\caption{}
\end{subfigure}\hspace{\fill}
\begin{subfigure}{0.49\linewidth}
\includegraphics[scale=0.175]{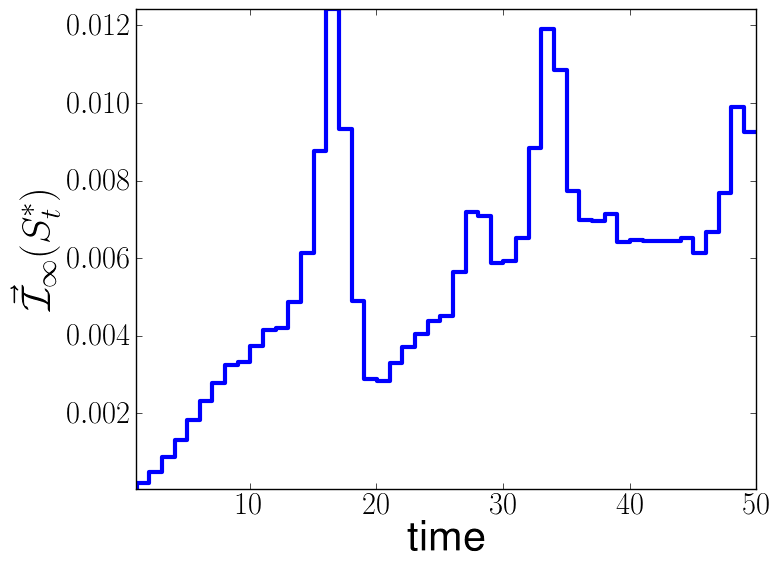}
\caption{}
\end{subfigure}\hspace{\fill}
\caption{Numerical simulation of the maze problem with $625$ cells. The sub figures (a), (c), and (e) correspond to $\beta=0.05$. They show the value landscape, the directed information landscape, and the directed information along the optimal path. The sub figures (b), (d), and (f) correspond to $\beta=0.005$. A phase-transition occurs between $\beta=0.05$ and $\beta=0.005$, where a typical behavior {\it "to pass through the narrow door"}, changes to another qualitatively different behavior {\it "to take a longer path without entering the narrow door"}. The peaks at the sub figures (e) and (f) correspond to the states with high information consumption. For instance, as seen at (e), the $31$st state along the optimal path (the entrance to the narrow door), corresponds to the maximal information rate that the agent should process to reach the goal state.}\label{expm1}
\end{center}
\vskip -0.2in
\end{figure}

Figure \ref{expm1} shows the value function, and the directed information for different values of parameter $\beta$. The optimal solutions to equation (\ref{eq:GREC2}) are characterized by {\it phase-transitions} - the qualitative difference in the typical behavior as a function of the value of parameter $\beta$. The optimal path is indicated by the solid white line from the start state, marked by $S$, to the goal state, marked by $G$.  

The low values of $\beta$ correspond to the high stochasticity in the system. The more stochastic the system is, the harder it is for the agent to affect the environment. Specifically, the agent prefers not to take the risk of going through the narrow door for $\beta=0.005$, as shown in Figure (\ref{expm1}-b). However, it does go through the narrow door for $\beta=0.05$, as shown in Figure (\ref{expm1}-a).   

The typical distribution of the directed information along the optimal path is shown in the bottom plots. The information is higher when the agent needs to move carefully, e.g., near the corners and/or through the narrow door. The solution is derived for the infinite time horizon, which enables us to explore the typical stationary policy of the agent.

As is seen in the sub figure (d), the agent needs different amounts of information at different times over the optimal path. For example, it must be more careful (needs more information), at the narrow hole in order to exit the compartment in the middle of the maze. This is the maximum of the minimal directed information from the states to the actions, which should be taken into consideration when choosing a processor for the agent.  

\section{Information from Agent to Environment}\label{sec:dualProb}
In this section we address the question of the maximal directed information from the agent to the environment, which is a dual question to the question of the maximal directed information from the environment to the agent. These two directed informations differ, because the directed information functional is asymmetric in general. We argue and demonstrate by numerical simulations that the maximal directed information between the agent and the environment is useful for acting optimally without an externally provided reward (unknown targets). 

\subsection{Background}
The information capacity of the channel from the agent's actions to the environment states was shown to be useful in solving various AI benchmark problems without providing an external reward/utility function, \cite{salge2014empowerment,klyubin2005empowerment, klyubin2005all, mohamed2015variational}. This capacity, known as {\it empowerment}, evaluates a maximal number of environment states that the agent can potentially achieve by its actions. 

The higher the value of empowerment is, the more potential states are reachable by actions. As in the empowerment method, when a goal state is not predefined but may occur in any state as in Chess, Go and other interactive games, the optimal strategy for the agent is to keep as many {\it reachable states} as possible. As was shown in \cite{klyubin2005all} Figure 5 and Figure 6, there is a qualitative similarity between the landscape of the empowerment and the landscape of the average shortest distance from a given state to any other state.    

In this paper we extend the empowerment method by considering in the feedback control policy by applying the directed information from the action trajectories to the environment trajectories. In the next section we describe the causal conditioning of a sequence of states of length $T$ on a sequence of actions of length $T$. We derive the Bellman-type recursion for the directed information from the action to the states, which together with the Bellman recursion from the states to the action in Eq.(\ref{eq:GREC2}) provide a complete picture of the bi-directional information exchange between the agent and the environment.

\subsubsection{Causal Conditioning}
We assume that at time $t$ the environmental state is $S_t$. Given $S_t$, the agent computes action $A_t$, and the environment moves to $S_{t+1}$, according to the given transition probability, $p(S_{t+1}\mid S_t, A_t)$. This interaction repeats $T-1$ cycles, driving the system to its final state, $S_T$, by a sequence of actions $A_1^{T-1}$. 

The joint probability between $A_{t}^{T-1}$ and $S_{t+1}^T$ conditioned on $S_t$ is given by $p(S_{t+1}^T, A_t^{T-1}\mid S_t)$. The causal dependency between the time series, $A_t^{T-1}$, and $S_t^T$ differs from the non-causal one in its conditional probability (for details see e.g., \cite{kramer1998directed, massey2005conservation, marko1973bidirectional,permuter2009finite}), as following:

{\small
\begin{align}
p(S_{t+1}^{T}\mid A_t^{T-1},S_t) =& \prod\limits_{i=t}^Tp(S_i\mid S^{i-1}_t, A_t^{T-1}),\\
p(S_{t+1}^{T}\parallelsum A_1^{T-1},S_t) =& \prod\limits_{i=t}^Tp(S_i\mid S^{i-1}_t, A_t^{i-1}).
\end{align}
} 

The join probability, $p(S_{i+1}^T, A_i^{T-1}\mid S_i)$, is given by:

{\small
\begin{align}
p(S^T_{t+1},& A^{T-1}_{t}\mid  S_t) =\nonumber\\
=& p(S_{t+1}^T\parallelsum A_{t}^{T-1}, S_t)p(A_{t}^{T-1}\parallelsum S_{t}^{T-1})\label{jointDistr1},\\
\intertext{{\normalsize which is decomposed alternatively according to Bayes' rule as:}}
=& p(A^{T-1}_{t}\mid  S^T_{t+1}, S_t)p(S^T_{t+1}\mid S_t).\label{jointDistr2}
\end{align}
}

\subsubsection{Directed Information}
The two alternative decompositions of the joint distribution, $p(S^T_{t+1}, A^{T-1}_{t}\mid  S_t)$, enable us to decompose the directed information as follows

{\small
\begin{align}
\mathcal{I}\Bigl[A_{t}^{T-1}\rightarrow& S_{t+1}^T \mid s_t\Bigr]=\label{eq:DIEMP2}\\
=&\left<\ln\frac{p(S_{t+1}^T\parallelsum A_{t}^{T-1}, S_t)}{p(S_{t+1}^T\mid S_t)}\right>_{p(S^T_{t+1}, A^{T-1}_{t}\mid  s_t)}\nonumber\\ 
{=}&\left<\ln\frac{\prod\limits_{j=t}^{T-1} p(A_{j}\mid S_{j}, S_{j+1})}{\prod\limits_{j=i}^{T-1}\pi(A_j\mid S_j)}  \right>_{p(S^T_{t+1}, A^{T-1}_{t}\mid  s_t)}\nonumber,
\end{align}
}  

which follows from the alternative decompositions of the joint distribution in (\ref{jointDistr1}) and (\ref{jointDistr2}):

{\small
\begin{align}
\frac{p(S_{t+1}^T\parallelsum A_{t}^{T-1}, S_t)}{p(S^T_{t+1}\mid S_t)}=\frac{p(A^{T-1}_{t}\mid  S^T_{t+1}, S_t)}{p(A_{t}^{T-1}\parallelsum S_{t}^{T-1})},
\end{align}
}

and the graph property: 

{\small
\begin{align}
1\le t<T : A_t \independent \left(A^{t-1}_1, A_{t+1}^{T-1}, S^{T-1}_1, S_{t+1}^T\right)\mid S_t, S_{t+1}. \nonumber
\end{align}
}

This decomposition leads to the Bellman equation for the directed information from the action trajectories to the state trajectories.
\subsection{Bellman Equation for $\vec{\mathcal{I}}[\mbox{Actions} \rightarrow \mbox{States}]$}
Separating the first term in (\ref{eq:DIEMP2}) from the rest we get the following recursion:

{\small
\begin{align}
	\mathcal{I}\left[A_{t}^{T-1}\rightarrow S_{t+1}^T \mid s_t\right]=&\mathcal{I}\left[A_t;S_{t+1}\mid S_{t}\right]\label{StateCondDirInfoRecur}\\
+&\left< \mathcal{I}\left[A_{t+1}^{T-1}\rightarrow S_{t+2}^T \mid S_{t+1}\right] \right>_{p(S_{t+1}\mid s_t)}.\nonumber
\end{align}
}

where $\mathcal{I}\left[A_t;S_{t+1}\mid S_{t}\right]$ is the mutual information between the $t$-th state and action, conditioned on the $(t+1)$-th state. The conditional directed information in (\ref{StateCondDirInfoRecur}) is a function of the state. To improve readability, the recursion in (\ref{StateCondDirInfoRecur}) may be represented concisely as follows, (where the right arrow in $\vec{\mathcal{I}}$ distinguishes the directed information from the ordinary/symmetric mutual information):

{\small
\begin{align}
	\vec{\mathcal{I}}(s_t)=\mathcal{I}\left[A_t;S_{t+1}\mid s_{t}\right]
+\left< \vec{\mathcal{I}}(S_{t+1})\right>_{p(S_{t+1}\mid s_t)}, \label{StateCondDirInfoRecurShort}
\end{align}
}

where the free distribution is the feedback policy distribution, $\pi(A\mid S)$. Assuming stationarity, the time index in (\ref{StateCondDirInfoRecurShort}) may be reduced to: 

{\small
\begin{align}
\vec{\mathcal{I}}(s)=\mathcal{I}\left[A;S'\mid s\right]
+\left< \vec{\mathcal{I}}(S^{'})\right>_{p(S^{'}\mid s)}. \label{StateCondDirInfoRecurShort2}
\end{align}
}
The maximum of $\vec{\mathcal{I}}(s)$ with regard to $\pi(A\mid S)$ gives the maximal number of states that the agent can achieve by its action. In the next section we provide the formal solution. 

\subsection{Optimal solution}
To find the maximum of the directed information in (\ref{StateCondDirInfoRecurShort2}), we apply the alternating maximization between the policy distribution, $\pi(A_i\mid S_i)$, and the {\it inverse-channel} distribution $p(A_i\mid S_i, S_{i+1})$ (which is conventionally denoted by $q(A_i\mid S_i,S_{i+1})$), as in the standard problem for finding the channel capacity, \cite{cover2012elements, blahut1972computation}. The optimal formal solution is given by the following set of self-consistent equations:

{\small
\begin{align}
&q^*(A\mid S, S^{'}) =\frac{p(S^{'}\mid S, A)\pi^*(A\mid S)}{  \sum\limits_{A} p(S^{'}\mid S, A)\pi^*(A\mid S) },\label{OptRev}\\
&\vec{\mathcal{I}}_{\pi^{*}}\left[s\right] = \mathcal{I}_{\pi^{*}}[A, S^{'}\mid S]+\left<\vec{\mathcal{I}}_{\pi^{*}}[S^{'}]\right>_{p(S^{'}\mid A, S)\pi^*(A\mid S)}\label{OptDI},\\
&\pi^*(A\mid S) =\label{OptPol}\\
&\qquad= \frac{\exp{\sum\limits_{S^{'}}p(S^{'}\mid S, A)\left( \ln q^*(A\mid S, S^{'})  + \vec{\mathcal{I}}^*[S^{'}]\right)}}{  \sum\limits_{A} \exp{\sum\limits_{S^{'}}p(S^{'}\mid S, A)\left( \ln q^*(A\mid S, S^{'})  +\vec{\mathcal{I}}^*[S^{'}]\right) }}\nonumber.
\end{align}
} 

This set of self-consistent equations is an extension of the common Blahut-Arimoto algorithm by the directed information of the future action and state trajectories, which appears in the exponent of $\pi^*(A\mid S)$ in Eq.\ref{OptPol}. A numerical solution is derived by iterating equations (\ref{OptRev}-\ref{OptPol}) until convergence. The typical convergence is shown by the numeric simulations, provided in the next section. 

\section{Numerical Simulations}\label{sec:UnknownGoal}
To demonstrate the optimal solution given by equations (\ref{OptRev}-\ref{OptPol}) we consider the problem of the optimal location of a firehouse in a city. We model the city by a grid world, where a fire engine is an agent moving across the city according to a given transition probability, $p(S_{t+1}\mid S_t, A_t)$. The goal is to locate the firehouse at a position where the firefighters can reach each block in the city in the fastest way on average. 

This is an example where the agent should be prepared optimally to reach any possible target state on average, rather than to be able to get to a particular state (or a number of states).

The naive solution is to compute the average shortest distance between each block in the city and all the other blocks, and to locate the firehouse on the block with the minimal average shortest distance. The algorithms for solving the all-pairs shortest path problem have a time complexity of $O(V^3)$ \cite{cormen2009introduction}, where $V$ is the number vertices in a graph.

\begin{figure}[t!]
\begin{center}
\begin{subfigure}{0.49\linewidth}
\includegraphics[width=\linewidth]{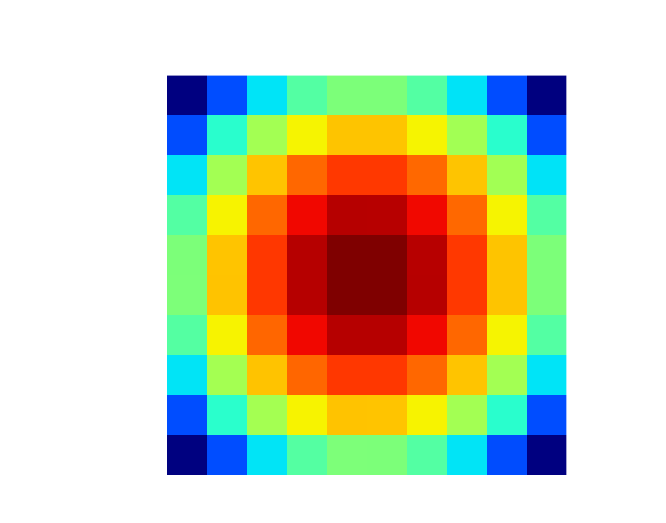}
\caption{}
\end{subfigure}
\begin{subfigure}{0.49\linewidth}
\includegraphics[width=\linewidth]{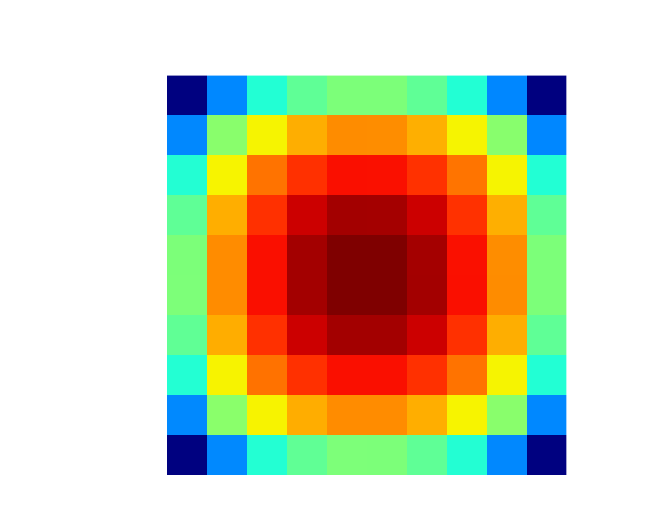}
\caption{}
\end{subfigure}

\begin{subfigure}{0.49\linewidth}
\includegraphics[width=\linewidth]{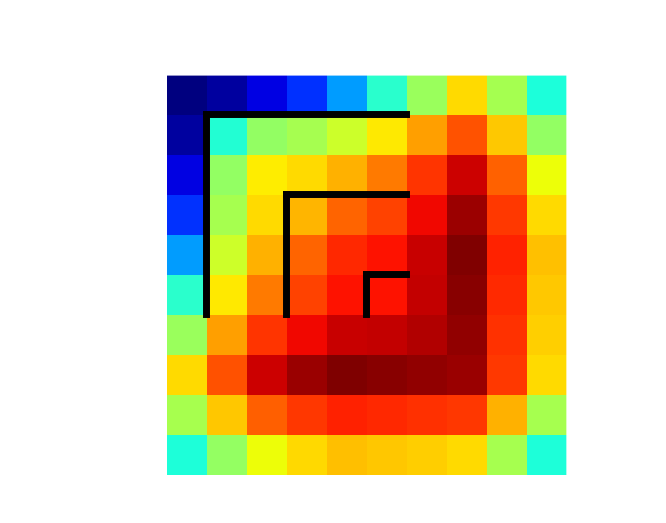}
\caption{}
\end{subfigure}
\begin{subfigure}{0.49\linewidth}
\includegraphics[width=\linewidth]{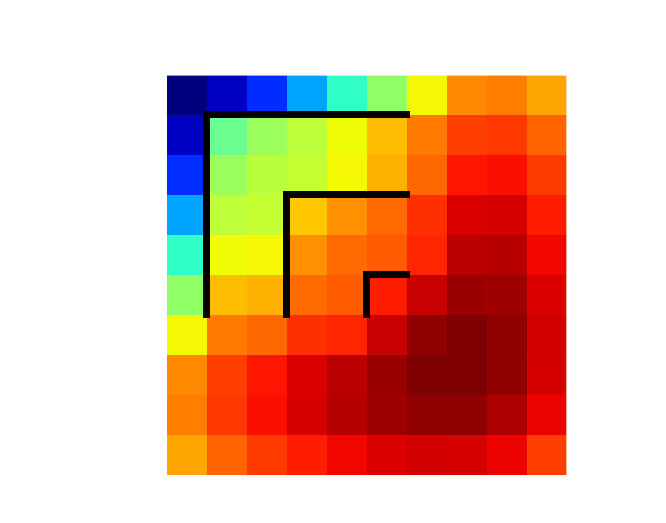}
\caption{}
\end{subfigure}\hspace{\fill}
\caption{Sub figures (a) and (c) show the exact values of the minimal average path from a given block to all the other blocks in the city. Sub figures (b) and (d) show the value of the maximal directed information from the agent to the environment in the long-time horizon on every block in the city. Black lines represent the fences in the city. Red corresponds to high values.}\label{fig:centrality} 
\end{center}
\end{figure}

The sub figures (\ref{fig:centrality}.a) and (\ref{fig:centrality}.c) show the exact values of the average shortest paths from a particular block to all the other blocks in two different cities, a city without any walls, and a city with perpendicular black walls, respectively. The sub figures (\ref{fig:centrality}.b) and (\ref{fig:centrality}.d) show the values of the directed information computed by iterating the self-consistent equation (\ref{OptRev}-\ref{OptPol}).  Red blocks denote high values, and blue blocks denote low values. 

The two approaches suggest qualitatively similar patterns on the location of the firehouse. The preferable locations are the brightest red blocks. In the case of the city without walls both solutions give the same block.  

\begin{figure}
  \begin{minipage}[c]{0.55\linewidth}
\includegraphics[scale=0.235]{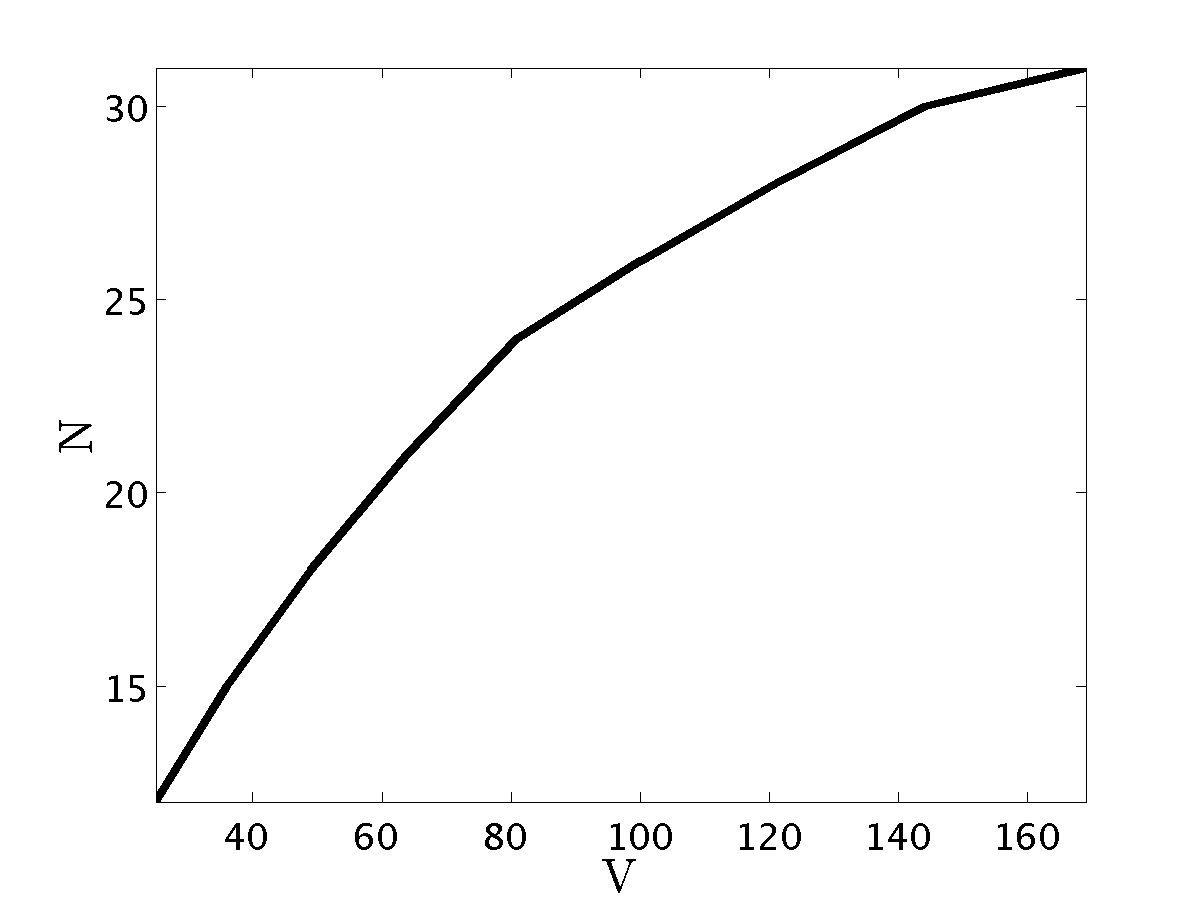}
  \end{minipage}
  \begin{minipage}[c]{0.4\linewidth}
\caption{Convergence of the Bellman recursion for the directed information. x-axis: number of blocks, y-axis: number of iterations until convergence.}\label{fig:convergance}
  \end{minipage}
\end{figure}

Figure (\ref{fig:convergance}) shows the convergence of the iteration of the self-consistent equations (\ref{OptRev}-\ref{OptPol}). The convergence is sub-linear in the number of blocks in the city, where each iteration consumes $O\Bigl(|\mathcal{S}|\times |\mathcal{A}|\Bigr)$ multiplications of real numbers, which is overall faster in order of magnitude than the Floyd/Dijkstra-type algorithms. Convergence is said to occur when the relative change in the directed information between two consequent iterations is smaller than tolerance, $\tau=10^{-6}$, $abs(\vec{\mathcal{I}}_{k+1} - \vec{\mathcal{I}}_{k})/\vec{\mathcal{I}}_{k}\le \tau$, where $\mathcal{I}_{k+1}$ is given by Eq.\ref{StateCondDirInfoRecurShort2} in the $k$-th iteration.  

\section{Conclusion}\label{sec:disc}
The feedback interaction between agent and the environment includes two asymmetric information channels. Directed information from the environment constraints the maximal expected reward obtained by the agent in the standard RL setting. Directed information is a functional of a joint probability distribution of agent's action trajectories and environmental state trajectories in a given time horizon. As a result, the value of the directed information depends on a particular time horizon. However, it is worth exploring the typical behavior of an agent in an infinite time horizon, when all transients events have settled down or become statistically insignificant. In this work we derive the typical behavior of an agent that strives to increase its expected reward under the limitations of directed information. This is shown to be possible through the unified Information-Value Bellman-type recursion, which we represent in this work. 

The second channel transfers directed information from the agent to the environment. It has been shown that the capacity of this channel is useful for solving various AI test bench problems, when the agent needs to be ready to solve a problem optimally on average for any possible target. We derived a Bellman recursion for this channel as well, which enables us to explore the typical behavior of the agent in an infinite time horizon.  

A direct continuation of this work would be to explore the entire action-perception cycle of the interaction between the intelligent agent and the environment when both information channels operate simultaneously.  

\subsection*{Acknowledgments} 
We thank Daniel Polani for many useful discussions. 
This work is partially supported by the Gatsby Charitable
Foundation, The Israel Science Foundation, and Intel ICRI-CI
center. 
 

\bibliography{icml2017_ST}
\bibliographystyle{icml2017}
\end{document}